\documentclass{article}
\usepackage{spconf,amsmath,graphicx,amsfonts,hyperref,dsfont}
\usepackage{xcolor}
\usepackage{booktabs}
\usepackage{arydshln}

\title{Signal Processing Grand Challenge 2023 -- e-Prevention: \\ Sleep Behavior as an Indicator of Relapses in Psychotic Patients}

\name{Kleanthis Avramidis\qquad Kranti Adsul\qquad Digbalay Bose\qquad Shrikanth Narayanan}
\address{Signal Analysis and Interpretation Lab, University of Southern California, Los Angeles, CA 90089}

\begin{document}
\maketitle

\begin{abstract}
This paper presents the approach and results of USC SAIL's submission to the Signal Processing Grand Challenge 2023 -- e-Prevention (Task 2), on detecting relapses in psychotic patients. Relapse prediction has proven to be challenging, primarily due to the heterogeneity of symptoms and responses to treatment between individuals. We address these challenges by investigating the use of sleep behavior features to estimate relapse days as outliers in an unsupervised machine learning setting. We extract informative features from human activity and heart rate data collected in the wild, and evaluate various combinations of feature types and time resolutions. We found that short-time sleep behavior features outperformed their awake counterparts and larger time intervals. Our submission was ranked 3rd in the Task's official leaderboard, demonstrating the potential of such features as an objective and non-invasive predictor of psychotic relapses.
\end{abstract}


\section{Introduction}
\label{sec:intro}

\vspace{-0.1cm}
Psychotic disorders are a category of mental illnesses that can significantly impact an individual's thoughts, emotions, and behavior. These disorders are characterized by a distorted perception of reality, which can manifest in the form of delusions, hallucinations, disordered thinking, and other cognitive impairments \cite{van2009schizophrenia}. Despite numerous related studies in neurophysiology \cite{mcgorry2014biomarkers}, effective biomarkers of psychotic episodes and relapses have not yet been established. This is due to the wide range of symptoms and variable treatment responses observed in patients \cite{gaebel1993early}. The potential of such biomarkers to timely diagnose or even prevent psychotic episodes is thus a prominent challenge in psychiatry.

Artificial Intelligence, and Machine Learning in particular, have emerged as promising tools in the search and identification of such possible markers, especilly those derived from biosensors \cite{aung2017sensing}. Nowadays, due to the adoption of wearables in everyday life, the potential of these studies is increasing. The e-Prevention project \cite{12345, panagiotou2022comparative} contributes in this direction by collecting features from physiological measures over the course of 6 months, on subjects in the psychotic spectrum.

The ICASSP Signal Processing Grand Challenge (SPGC) 2023 aimed to provide resources and algorithms to advance relapse days detection in the bio-sensing context. Our challenge submission, focused on sleep-related sensing, was ranked 3rd in the official leaderboard. Sleep disorders have been identified to correlate with psychotic episodes \cite{kaskie2017schizophrenia}. In this paper we present our approach, including feature extraction schemes over multiple temporal resolutions and during sleep and awake periods. Our results using an Isolation Forest algorithm for outlier detection, discriminated relapse from normal days with an AUC of 60.5\% on the test set. \vspace{-0.11cm}

\section{Data Processing}
\label{sec:method} \vspace{-0.2cm}

\subsection{Dataset}

As part of the e-Prevention project, 37 patients on the psychotic spectrum were recruited and provided with a Samsung Gear S3 smartwatch. Through the wearable, the researchers collected measures of linear and angular acceleration (20Hz), RR peak intervals derived via photoplethysmography (5Hz), sleeping schedule, and step count, for a total monitoring period of up to 2.5 years. Clinicians then annotated patients’ relapse days in cooperation with their physicians. The challenge provided us with a subset of 6-month daily data for 10 patients. The data come in 3 splits, namely train split, that contains only non-relapse days, validation split, categorized in relapse and non-relapse days, and unlabeled test split. \vspace{-0.15cm}

\subsection{Feature Extraction}

In the pre-processing stage, we removed outliers and imputed missing values at a range of 1 hour with the use of the Hampel method \cite{Liu2004OnlineOD}. For feature extraction, we derived several types of features from 5-minute intervals of the processed time-series, which were then aggregated at various resolutions. Specifically, we computed the normalized energy of the accelerometer and gyroscope measurements to account for changes in movement and activity levels. We also extracted the mean heart rate (BPM) and heart rate variability (HRV) from the RR intervals. Initially, the power spectral density of these features was estimated using Welch's method, after which the low (LF) and high frequency (HF) bands, along with their respective fractions, were isolated \cite{12345}. Daily sinusoidal encoding was employed for the timestamp feature.

Regarding the step count data, we incorporated the provided features and computed additionally step size and speed by applying a time-to-seconds conversion on the start and end times of the steps. The data are then distributed over 5-minute intervals by summing up the step counts and taking the mean of the distance, calories, step size, and speed. \vspace{-0.15cm}

\section{Experimental Setup}
\vspace{-0.1cm}

\begin{table}
\centering
  \begin{tabular}{lccc}
    \toprule
    \textbf{Features}&\textbf{5-minute}&\textbf{60-minute}&\textbf{Aggregate}\\
    \midrule
    Sleep &  63.9 \% & 59.1 \% & 61.7 \% \\
    Awake & 59.4 \% & 52.5 \% & 55.5 \% \\
    Aggregate & 61.0 \% & 54.0 \% & 57.8 \% \\
    \midrule
    Sleep + Step & \textbf{64.5} \% & 59.4 \% & 61.8 \% \\
    Awake + Step & 60.1 \% & 49.0 \% & 54.9 \% \\
    Aggregate & 61.2 \% & 53.9 \% & 60.1 \% \\
    \bottomrule
\end{tabular}
\caption{Results summary for different experiment versions.}
\label{tab:results_main}
\vspace{-0.4cm}
\end{table}

The feature vectors outlined above are formed by considering the mean and standard deviation within non-overlapping 5-minute intervals, also aggregated to 1-hour intervals. After extracting the features for each participant, we standardize them to unit norm and concatenate them to run subject-agnostic trials. We compare using only data during sleep (or awake) and with or without step count information.

\textbf{Models:} The problem in hand is essentially a novelty detection task, wherein the objective is to identify whether a given sample is an outlier in the absence of any outlier data in the training set. To address this task, we evaluated a range of tree-based and clustering-based algorithms, and we selected Isolation Forest \cite{liu2008isolation} for further experiments. Isolation Forest is a tree-based ensemble method that works by randomly selecting a feature and then randomly splitting between its extreme values. The number of splits required to isolate a sample is then used as a measure of normality, since random partitioning produces shorter paths for anomalies. This measure is averaged through a forest of such random trees.

\textbf{Evaluation:} Given that the task involved a ranking evaluation of relapse days, we assessed the model's performance and statistical significance by computing ROC-AUC and PR-AUC ranking scores. The ultimate reported metric is their harmonic mean, computed across subjects. In the absence of sleep-oriented features, we calculate the AUC score on the awake data and standardize based on the optimal relapse threshold of the sleep-related predictions. \vspace{-0.1cm}

\section{Results}
\label{sec:exp}

\vspace{-0.1cm}
Table~\ref{tab:results_main} presents the results of all versions of our experiments, with each score indicating the aggregate AUC metric, which is calculated for the samples of the validation set that include the respective features. Our analysis revealed that 5-minute-level features were more discriminative than 1-hour resolution features, with a 5--7\% increase in performance during sleep and awake periods, respectively. And this comes in spite of the less data samples; the overall computation time was three times faster with the sleep features compared to the features from throughout the day. Additionally, features extracted during sleep were more informative than those obtained throughout the day, resulting in 5--10\% increases in AUC score. Concatenating step count information also improved the derived performance, only slightly though. Our experimentation suggests that the optimal parameter setting is 5-minute data with step features during sleep, standardized to unit norm. \vspace{-0.12cm}

\section{Conclusion}
\label{sec:conc}

\vspace{-0.1cm}
This paper presents USC SAIL's approach and results in the Signal Processing Grand Challenge 2023 for detecting relapses in psychotic patients. We investigated the use of sleep activity and heart rate features and evaluated various combinations of feature types and time resolutions. Our results show that short-time sleep behavior features outperformed their awake counterparts and larger time intervals, scoring 64.5\% on the validation set and 60.5\% on the test set. \vspace{-0.1cm}

\section{Acknowledgements}

\vspace{-0.1cm}
We want to thank for their contribution the USC SAIL's members that competed in SPGC Task 1: Anfeng Xu, and Tiantian Feng,  as well as our collaborators from the Department of Artificial Intelligence, Wroclaw University of Science and Technology, Stanislav Saganowski, and Bartosz Per.

\ninept
\bibliographystyle{IEEEbib}
\bibliography{refs}

\end{document}